\documentclass[11pt]{article}

\usepackage[utf8]{inputenc}
\usepackage[T1]{fontenc}
\usepackage{lmodern}
\usepackage{geometry}
\usepackage{microtype}
\usepackage{graphicx}
\usepackage{subcaption}
\usepackage{multirow}
\usepackage{amsmath,amssymb,amsfonts}
\usepackage{amsthm}
\usepackage{mathrsfs}
\usepackage[title]{appendix}
\usepackage[table]{xcolor}
\usepackage{textcomp}
\usepackage{manyfoot}
\usepackage{booktabs}
\usepackage{tabularx}
\usepackage{threeparttable}
\usepackage{array}
\usepackage{algorithm}
\usepackage{algorithmicx}
\usepackage{algpseudocode}
\usepackage{listings}
\usepackage{float}
\usepackage{authblk}
\usepackage[numbers,sort&compress]{natbib}
\usepackage{hyperref}
\hypersetup{colorlinks=true,linkcolor=darkblue,citecolor=darkblue,urlcolor=darkblue}

\definecolor{darkblue}{rgb}{0.0,0.0,0.55}

\newcommand{\bmhead}[1]{\par\medskip\noindent\textbf{#1}\par\smallskip}
\newcommand{\backmatter}{}

\DeclareUnicodeCharacter{2018}{`}
\DeclareUnicodeCharacter{2019}{'}
\DeclareUnicodeCharacter{201C}{``}
\DeclareUnicodeCharacter{201D}{''}
\DeclareUnicodeCharacter{2013}{--}
\DeclareUnicodeCharacter{2014}{---}

\theoremstyle{plain}

\theoremstyle{remark}

\theoremstyle{definition}

\begin{document}

\title{The Evolution of Digital Twins from Reactive to Agentic Systems}

\author[1]{Omer San\thanks{Corresponding author. Email: \texttt{osan@utk.edu}}}
\author[2]{Adil Rasheed}
\author[3]{Eda Bozdemir}
\author[4,5]{Jun Deng}

\affil[1]{Department of Mechanical and Aerospace Engineering, University of Tennessee, Knoxville, TN 37966, USA}
\affil[2]{Department of Engineering Cybernetics, Norwegian University of Science and Technology, Trondheim, NO-7465, Norway}
\affil[3]{Department of Pathology, Yale University, New Haven, CT 06510, USA}
\affil[4]{Department of Therapeutic Radiology, Yale University, New Haven, CT 06510, USA}
\affil[5]{Department of Biomedical Informatics and Data Science, Yale University, New Haven, CT 06510, USA}

\date{}

\maketitle

\emph{Standfirst:}
Digital twins are evolving into self-learning, autonomous systems that link models, data, and human interaction. Realizing their full potential depends on interoperability, standardization, and the integration of artificial intelligence and advanced computational reasoning across sectors.

\section*{Introduction} \vspace{-5pt} 
As often discussed in the context of complex systems and chaos theory, and sometimes attributed to John von Neumann or Edward Lorenz, some systems are hard to predict but easy to control, while others are easy to predict yet hard to control. Financial markets, for example, often defy long-term prediction, but small and timely policy interventions, such as adjustments in central bank interest rates, can steer their behavior in measurable ways. In contrast, large-scale engineered infrastructures, like power grids, can often be modeled and forecasted with high precision, yet they remain difficult to control in real time due to delays, uncertainty, and physical constraints. These contrasts illustrate a broader truth: every complex system presents its own compromise between predictability, controllability, and the scope for effective intervention.

Digital twins (DTs) are emerging as a unifying framework to navigate this complexity. By coupling models, data, and intelligence in a continuously learning loop between physical and virtual representations, digital twins enable both improved prediction and adaptive control. Yet not every DT faces the same challenges, nor do they all rely on identical enabling technologies. The degree to which a digital twin can sense, reason, and act depends on its underlying data, models, control strategies, integration capabilities, and trust metrics.


In this Comment, we focus on the capability levels that differentiate them. Building on emerging classification schemes adopted by standardization bodies and industry leaders \citep{Stadtmann2023}, we discuss how these capability layers, ranging from standalone digital twins to fully autonomous systems, offer a structured way to evaluate maturity, identify gaps, and guide future research and deployment.

\section*{The Expanding Vision of Digital Twins} \vspace{-5pt}
Starting in the aerospace and manufacturing sectors, the concept of the DT was originally conceived as a virtual counterpart of a physical system that mirrors its behavior in real time \citep{Tao2018,Ferrari2024}, and over the past decade it has rapidly expanded into fields such as medicine, where patient-specific twins in cardiology and oncology enable personalized monitoring, prediction, and intervention \citep{Willcox2023,Katsoulakis2024}. Over the past decade, the DT vision has expanded beyond simple geometric or physical modeling to encompass broader notions of intelligence, autonomy, and continuous learning. Beyond traditional modeling and simulation, the central premise is bidirectional data exchange between the physical asset and its virtual representation, allowing the twin not only to reflect but also to influence the behavior of the physical asset \citep{Rasheed2020}.
 
Despite its popularity, “digital twin” has become an umbrella term encompassing diverse interpretations across disciplines, shaped by differing technological narratives and the distinct priorities of various communities. Some implementations refer merely to data dashboards or isolated simulations without closed-loop feedback or adaptive learning. In many cases, these rely on models or simulation tools designed to run on cloud or edge computing infrastructures, providing scalability but not necessarily dynamic coupling with the physical asset. Such systems are better described as digital shadows or digital models, rather than true twins, as they lack the continuous, bidirectional flow of information and control that defines a genuine digital twin. The essence of a DT lies in its \emph{bidirectional flow of information}, from the physical asset to update the virtual model, and from the virtual model back to optimize, predict, or control the physical system \citep{Kapteyn2021}.

With the rapid progress in artificial intelligence (AI) and agentic systems, a key advantage of digital twins lies in their capacity for automated knowledge generation. Recent advances in transformer-based architectures have shown that complex reasoning can emerge from modular, task-specialized agents capable of collaboration through shared representations. Extending this paradigm to digital twins enables distributed intelligence, in which multiple agents operate on distinct subtasks such as perception, control, or optimization, yet collectively enhance the twin’s adaptability and decision-making capacity. By continuously integrating sensor data, simulation outputs, and historical records, a twin can uncover patterns and insights that remain hidden in conventional analyses. This capability allows researchers to generate new hypotheses, identify subtle correlations, and refine predictive models without extensive manual intervention, thereby effectively accelerating the discovery cycle across complex systems.

Such acceleration is especially valuable for critical assets, whether defense systems, industrial machinery, or biomedical devices, but often pose challenges for direct experimentation due to safety, cost, ethical considerations, or limited accessibility. Digital twins provide a secure and efficient framework for learning from such assets, enabling models to be trained and refined without endangering the system or its operators. They also address key practical barriers such as long turn-around times and restricted laboratory access, allowing on-demand experimentation where alternative scenarios can be explored computationally before committing to physical tests. In doing so, digital twins maximize the utilization of laboratory resources, support informed decision-making, and enable optimized control strategies. Physical experiments can be queued, scheduled, or pre-screened through their virtual counterparts, ensuring efficient equipment use and, in public-facing domains, enhanced stakeholder engagement through transparent and interactive simulations.

Finally, digital twins support the systematic archival of data and knowledge, preserving not only results but also the \emph{knowledge depth} accumulated throughout experimentation and analysis. By maintaining structured, searchable records of both physical and virtual experiments, they ensure that insights are retained, traceable, and reusable across future studies. This archival capability underpins long-term learning, facilitates regulatory compliance, and strengthens the overall reliability and accountability of the systems under consideration.

\section*{What Is a Digital Twin?} \vspace{-5pt}

\begin{figure}[h!]
    \centering
    \includegraphics[width=1.0\textwidth]{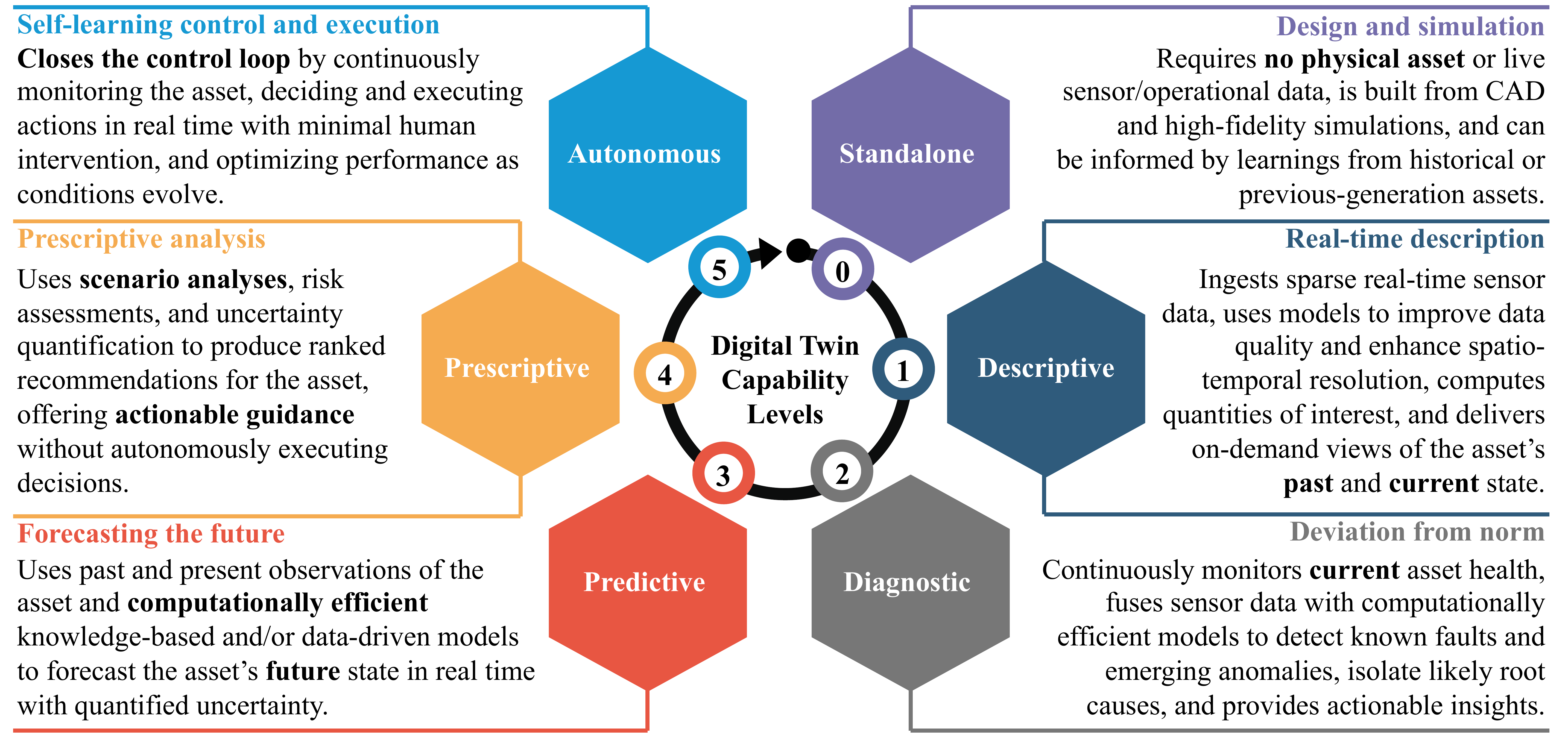}
    \caption{Capability-based classification of digital twins. The hierarchy progresses from standalone representations and descriptive monitoring to diagnostic, predictive, prescriptive, and fully autonomous systems. Each level reflects an increasing degree of integration, intelligence, and decision-making autonomy \citep{San2021}.}
    \label{fig:dt}
\end{figure}

A digital twin is defined as a virtual representation of a physical asset or a process enabled through data and simulators for real-time prediction, optimization,
monitoring, control, and informed decision-making \cite{Rasheed2020}. As shown in Figure~\ref{fig:dt}, a capability-based classification provides a consistent framework for understanding DTs. Capability levels define a hierarchy from simple representations to fully autonomous, agentic systems, highlighting the progression of functionality, intelligence, and automation. The scale ranges from 0 to 5, with each level building upon the previous, enabling systematic assessment, development planning, and cross-domain comparison. Importantly, DTs are objective-oriented: depending on the system and operational goals, a descriptive or prescriptive twin may already provide substantial value, whereas predictive or autonomous twins may be necessary for more complex applications. This layered approach also helps contextualize enabling technologies and facilitates standardization.

\emph{Standalone digital twins} represent an asset independently of its real-world counterpart, often before the physical system exists. For example, Computer-Aided Design (CAD) models capture the geometry and configuration of an asset, while high-fidelity simulation tools reproduce its expected physical behavior. Both provide valuable design insights but operate without real-time operational data, and therefore remain digital representations rather than true twins.

\emph{Descriptive digital twins} integrate real-time sensor streams with digital representations to maintain an up-to-date view of the asset, supporting continuous monitoring and situational awareness. A descriptive twin ingests sparse, real-time sensor data, uses models to improve data quality and enhance spatio-temporal resolution, and computes key quantities of interest to infer otherwise unmeasured states. By fusing data and models in this way, it can deliver on-demand visualizations of the asset’s past and present behavior, providing an accurate, evolving picture of system performance without yet enabling higher-level diagnostics or prediction.

\emph{Diagnostic digital twins} extend descriptive capabilities by fusing sensor data with computationally efficient models to continuously assess asset health and identify deviations from expected behavior. They monitor real-time performance, detect known faults as well as emerging anomalies, and isolate likely root causes before failures propagate. By comparing measured responses with model-based predictions, diagnostic twins can distinguish normal operational variability from abnormal trends. This enables early intervention, informed maintenance planning, and the generation of actionable insights that enhance system reliability and resilience.

\emph{Predictive digital twins} use data and models to forecast future system states in real time, enabling proactive planning and prognostics. They integrate past and present observations of the asset with computationally efficient, physics-based, and data-driven models to anticipate its future behavior with quantified uncertainty. By continuously assimilating sensor data and simulation outputs, predictive twins can update their forecasts as new information becomes available, allowing for adaptive and resilient decision-making. This capability supports condition-based maintenance, risk assessment, and operational optimization, transforming digital twins into forward-looking instruments of prediction and preparedness.

\emph{Prescriptive digital twins} extend predictive capabilities by integrating scenario analysis, risk assessment, and uncertainty quantification to generate ranked, actionable recommendations. They evaluate alternative operating strategies and explore “what-if” scenarios to guide decision-makers toward optimal responses under varying conditions. Although prescriptive twins do not execute actions autonomously, they bridge the gap between prediction and control by translating analytical insights into informed, evidence-based guidance for scientists, clinicians, policymakers, and technical practitioners across domains.

\emph{Autonomous digital twins}, in contrast, close the loop by continuously monitoring system behavior, making context-aware decisions, and executing control actions in real time with minimal human supervision. At this stage, the bidirectional flow of data, insights and control actions between the physical asset and its digital counterpart is fully realized. In meteorology, for instance, one might argue that modern weather forecasting models represent some of the most sophisticated digital twins. Nevertheless, they are better described as predictive rather than autonomous twins, since they excel at forecasting but do not act to modify the weather through feedback. 

\begin{table*}[!htbp]
\centering
\caption{Examples of digital twin capability levels in meteorology and medicine.}
\label{tab:dt-capability}
\footnotesize
\begin{threeparttable}
\begin{tabularx}{\textwidth}{@{}>{\raggedright\arraybackslash}p{2.0cm}X X@{}}
\toprule
\textbf{Level} & \textbf{Meteorology} & \textbf{Medicine} \\
\midrule
0. Standalone  &
Static CAD model of terrain or atmospheric geometry without data assimilation. &
Anatomical CAD model without physiological data coupling or real-time measurements. \\

\addlinespace[4pt]
1. Descriptive &
Multimodal observations and simulation model outputs fused and visualized in realtime. &
Wearable data, health records / history, combined to represent the patient’s current health profile. \\

\addlinespace[4pt]
2. Diagnostic &
Detection of abnormal atmospheric dynamics, e.g., early cyclone formation. &
Identification of irregular vital signs such as abnormal blood pressure or heart rhythm. \\

\addlinespace[4pt]
3. Predictive &
Numerical weather prediction models forecasting future atmospheric conditions. &
Health-trajectory forecasts under current medication or treatment plans. \\

\addlinespace[4pt]
4. Prescriptive  &
Model-based testing of emission scenarios and policy-driven mitigation strategies. &
Simulation of alternative therapies or dosages to optimize expected health outcomes. \\

\addlinespace[4pt]
5. Autonomous &
Active weather modification or automated microclimate regulation (e.g., targeted cloud seeding). &
Closed-loop insulin delivery systems automatically regulating glucose levels. \\
\bottomrule
\end{tabularx}
\end{threeparttable}
\end{table*}

This classification highlights that DTs are designed to deliver value according to their operational objectives. Table~\ref{tab:dt-capability} illustrates two examples highlighting how increasing integration of data, analytics, and autonomy transforms domain-specific applications from static representations to self-governing systems. Early-stage design phases may benefit most from standalone or descriptive twins, which emphasize geometry definition, visualization, and conceptual exploration. In contrast, predictive, prescriptive or autonomous twins become essential during operational and optimization stages, where feedback, adaptation, and decision support are critical. CAD models or high-fidelity simulations, while sometimes labeled as digital twins, fall under the standalone category. As digital twins mature, these models evolve toward interactive, multimodal, heterogeneous environments. Emerging approaches such as surrogate modeling, scientific machine learning, and reduced-order modeling are accelerating this transition by enabling real-time performance, uncertainty quantification, and the continuous integration of experimental or operational data. These advances often serve as transforming engines, converting computationally intensive high-fidelity simulations into efficient, on-demand tools. This transformation typically incurs an offline training cost but yields substantial online speed and adaptability, allowing digital twins to operate interactively and at scale. In doing so, they extend legacy modeling and simulation capabilities while preserving relevance and interoperability across all digital twin capability levels. The extent to which these benefits are realized, however, can differ considerably between disciplines.

Across sectors, the maturity of digital twin implementation varies widely along the capability spectrum, largely reflecting the intrinsic data density and control architecture of each field. In the process and manufacturing industries, digital twins have reached semi-autonomous functionality, although their operation is typically enabled through rule-based logic or traditional \emph{proportional–integral–derivative} control schemes rather than fully adaptive or cognitively autonomous architectures. Conversely, the construction and architecture sector often treats building information models as largely standalone tools focused on logistics, scheduling, and remote visits, although a clear trend toward systems that optimize energy use and indoor comfort is emerging. In engineering and infrastructure, including aerospace and energy, the focus remains primarily diagnostic to predictive, coupling high-fidelity simulations with sophisticated data assimilation and uncertainty quantification techniques. Weather and climate modeling represents one of the most mature large-scale implementations, leveraging decades of standardized data exchange and coordinated global assimilation networks; in specialized tasks such as adaptive observations and ensemble prediction, these systems achieve high levels of predictive capability. Biomedical and healthcare applications, constrained by challenges in data availability, interoperability, and patient-specific longitudinal records, are largely confined to descriptive and diagnostic stages, despite rapid progress in cardiology and oncology. Collectively, these distinctions underscore how a layered taxonomy of digital twin capabilities provides a comparative lens across domains.

Overall, the layered framework fosters a multidisciplinary understanding of DTs, emphasizing that modeling, control, and simulation are not ends in themselves. Rather, digital twins function as communicative, objective-driven systems that integrate data, models, AI, and human interaction to generate actionable insights and value across diverse applications.

\section*{Data, Models, and Control as the Backbone of Digital Twins} \vspace{-5pt}

The foundation of an effective DT rests on the seamless integration of data, models, and control. These three core components form a dynamic feedback system in which data informs models, models generate insight and prediction, and control mechanisms act upon the physical asset to achieve desired outcomes. A well-designed twin therefore operates not as a passive replica but as an intelligent intermediary between observation, understanding, and action.

Data remains the essential substrate of this ecosystem. The challenge lies not in acquisition but in transforming raw measurements into structured, actionable knowledge. This requires integrating heterogeneous and multimodal sources, including sensor streams, physics-based models, domain knowledge, and user interactions, within a unified computational framework \citep{Tao2018, Rasheed2020, Kapteyn2021}. As digital twin systems grow in complexity, spanning sub-models, communication interfaces, data storage and generation, data standardization and ontology development have emerged as critical enablers. Consistent digital ontologies and shared taxonomies allow diverse models and data sources to interoperate seamlessly, ensuring interoperability, scalability, and long-term sustainability. Such integration supports real-time synchronization and underpins trust, traceability, and reusability across the entire digital twin ecosystem. 

In meteorology, for example, data standardization and interoperability have reached a high level of maturity, as the field depends on the seamless exchange of massive datasets between observational, modeling, and assimilation centers. Global standards for data formats and metadata enable effective coupling between measurement systems and predictive models, supporting real-time forecasting and longer-horizon, multi-scale climate analysis. 
In medical applications, by contrast, achieving comparable levels of data availability and standardization remains difficult, constrained by ethical oversight and the intrinsic heterogeneity of healthcare systems. Longitudinal datasets spanning extended time horizons are essential for constructing accurate and personalized models that capture the dynamic evolution of individual patients. Consequently, some of the earliest applications of digital twin technologies have emerged in cardiology, oncology, and endocrinology, where continuous monitoring and individualized modeling are central \citep{Laubenbacher2024}.

Models form the second backbone, providing a computational abstraction of the underlying system. They may include physics-based models, data-driven models, or hybrid combinations that harness the strengths of both approaches to achieve accuracy, efficiency, and generalization. Such integration enables digital twins to represent both deterministic dynamics governed by physical laws and stochastic variability arising from measurement noise, model uncertainty, or unresolved physics. Recent advances are shifting focus from static model construction toward adaptive modeling frameworks that evolve with new data, enabling continuous recalibration, uncertainty quantification, and cross-regime generalization. These developments allow digital twins to operate efficiently while retaining interpretability, ensuring that models remain reliable as systems age, degrade, or encounter new operational conditions. Together, data and models allow digital twins to evolve from descriptive and diagnostic tools toward predictive and prescriptive systems capable of forecasting, optimization, and adaptive control.

Control completes the feedback loop. Through dynamic coupling between the physical and digital layers, control architectures allow real-time decision-making, intervention, and self-optimization. Intelligent control systems embedded within digital twins can autonomously adjust parameters, reconfigure operational modes, or schedule experiments in response to system feedback. Designing such controllers, however, remains resource-intensive and requires iterative tuning as systems evolve, sensors degrade, or objectives shift. By embedding adaptive and learning-based control frameworks, digital twins can overcome these limitations, continuously refining control strategies and assimilating new operational knowledge. This interplay between perception, modeling, and actuation transforms digital twins into active participants in complex environments rather than mere observers.

At the convergence of these pillars, large language models (LLMs) are emerging as powerful enablers. Beyond their language capabilities, LLMs serve as semantic mediators linking data streams, model outputs, and control decisions within a unified framework \citep{Rasheed2025}. They enhance the operability and visualization of digital twins while augmenting perceptual and cognitive capacities, strengthening interpretability and situational awareness across interconnected systems. LLMs also bridge human intent and machine execution. They can express control objectives in natural language, interpret heterogeneous inputs such as telemetry, documentation, and tabular data, and, when coupled with databases, simulators, or learned predictors, evaluate alternative actions without requiring deep control expertise. Continuously updated with new knowledge, they cultivate knowledge depth within the twin by assimilating evolving domain practices and standards. This adaptability aligns with the data-centric nature of digital twins, where process information is persistently refreshed for informed and autonomous decision-making.

\section*{Interfaces for Digital Twins} \vspace{-5pt}

Interfaces form the connective tissue of a digital twin system. They mediate exchanges between the physical asset and its virtual counterpart, among subcomponents within the twin, and between the twin and human operators. Well-designed interfaces ensure that data, models, and control actions remain coherent, synchronized, and interpretable across scales. They also organize layered representations and visual abstractions that connect raw sensor data to high-level insights, allowing users to navigate between physical, computational, and cognitive layers of the system. Collectively, interfaces form the connective architecture that links sensing, computation, and decision-making in a unified framework.

Within this broader context, augmented reality (AR) has emerged as a critical interface that enhances the interpretability and interactivity of digital twins. Through immersive visualization, AR bridges the cognitive gap between complex simulation data and human decision-making \citep{Billinghurst2015}. Engineers, operators, and scientists can visualize latent physical quantities, inspect operational states in real time, and even manipulate parameters in an interactive 3D environment.  

By overlaying virtual information onto real-world environments, AR allows humans to perceive the state of the twin and the asset as a unified system. This mixed-reality interaction is not merely a display mechanism; it enables embodied engagement, where human intuition, sensor data, and model intelligence coexist in a shared context. At the subsystem level, interfaces also coordinate information flow among digital modules such as solvers, databases, and learning agents, ensuring consistent coupling and modular scalability. When coupled with machine learning models and physics-informed simulators, AR-enhanced twins provide intuitive ways to collaborate with AI systems.

Currently, AR interfaces primarily rely on interaction through speech, touch, and hand gestures, enabling users to query the twin, adjust parameters, and explore simulations directly within the augmented environment. Advances in wearable technologies, such as smart glasses integrated with electromyography sensors, are enabling richer and more precise control, allowing users to interact with twins through subtle muscle movements or hand posture recognition.  

Looking forward, near-future AR systems are likely to incorporate electroencephalography-based brain-computer interfaces and multimodal biosensing, offering direct non-invasive neural communication with digital twins. Such developments could allow operators to influence simulations, monitor physiological responses, or guide decision-making without physical gestures, further shrinking the gap between human intent and digital action. The integration of smart bands, eye-tracking, and haptic feedback into AR wearables will make these interactions seamless, natural, and context-aware, providing a continuous loop of perception, cognition, and control.

\section*{From Reactive to Agentic Twins} \vspace{-5pt}

Traditional DTs have been largely \emph{reactive} systems: they collect sensor data, perform simulations, and provide insights that help humans make better decisions. In these systems, actions are driven by external events or stimuli rather than self-directed processes. The twin responds to incoming data but does not initiate actions on its own, as the feedback loop is completed externally through supervisory control or human decision-making. By contrast, the emerging generation of digital twins is becoming agentic, meaning that the twin can formulate goals, plan actions, and interact with its environment with a degree of autonomy \citep{Hughes2025, Antonesi2025}. 

Agentic behavior aligns with the higher capability levels, where autonomous twins close the loop through self-directed reasoning and control. However, the reactive–agentic distinction is not a second taxonomy and should not be interpreted as mapping one-to-one onto the capability scale. Instead, it is a behavioral dimension that can manifest at different levels of capability. A twin may operate reactively even at advanced levels if its role is limited to responding to events or executing externally defined policies, while elements of agentic behavior can appear at intermediate stages when subsystems begin to set goals, adapt models, or plan actions within constrained scopes.

With this perspective in mind, agentic AI brings a layer of cognitive capability that fuses perception, reasoning, and action within the twin. It endows digital twins with a form of operational self-awareness, allowing them to assess system states, anticipate outcomes, and adjust strategies accordingly. Rather than passively representing the physical system, these agentic twins can autonomously explore design alternatives, plan interventions, and negotiate trade-offs under uncertainty. Through advances in reinforcement learning, causal modeling, and multimodal perception, these systems now realize the long-envisioned concept of “cognitive twins” proposed in the context of Industry 4.0 \citep{Zheng2022}, translating it into concrete and operational AI frameworks.
 
LLMs play a transformative role in this shift. They provide a common language for interpreting unstructured data, formulating goals, and communicating with human operators. However, as Handler et al. have highlighted, their fragility and tendency toward hallucination, particularly in high-stakes domains such as medicine where decisions may carry life-or-death consequences, make such errors as dangerous as outright failures \citep{Handler2025}. When integrated with physics-based models and domain-specific simulators, LLM-driven agents can reason about scenarios, explain outcomes, and adjust strategies based on human feedback.  

Agentic DTs are not merely digital surrogates but cognitive collaborators that learn from experience, reason through natural language, and act through autonomous control. They operate through mission-oriented coordination, where specialized agents oversee sensing, modeling, calibration, and control, collectively maintaining adaptability and resilience. They embody a shift from tools that analyze to systems that perceive, decide, and evolve in partnership with humans. By integrating adaptive learning, goal-directed reasoning, and contextual awareness, these twins move beyond passive representation toward self-improving, purpose-driven intelligence. This transition marks a conceptual bridge between current digital-twin technology and emerging forms of integrated cognitive systems, where physical and digital modes of reasoning begin to converge.

\section*{Challenges and Outlook} \vspace{-5pt}

Despite the conceptual elegance of this integration, realizing such systems presents formidable challenges. First, data governance, interoperability, and regulatory compliance remain unresolved. Physical systems operate across fragmented platforms with heterogeneous standards for sensing, storage, and semantics, while safety requirements and industry regulations impose additional constraints on how digital twins can be designed, deployed, and updated. Second, trust, explainability, and cybersecurity must evolve alongside capability. Agentic twins capable of autonomous decision-making demand mechanisms for accountability, auditability, ethical alignment, and robust protection against cyber threats. Third, computational sustainability and energy efficiency are critical concerns, as training and deploying large models within real-time digital twin frameworks can incur significant energy and hardware costs.

To address these challenges in practical settings, several methodological advances are essential. These include devising algorithms that harmonize and aggregate heterogeneous multimodal and multiscale data for effective fusion within DTs; investigating under-sampling uncertainty in systems with large spatiotemporal variability to clarify the limits of fidelity and reliability; and developing hybrid models that combine mechanistic formulations with data-driven representations to enable DTs to operate seamlessly across multiple scales. Parallel efforts are needed to build uncertainty quantification frameworks that capture uncertainties arising from both mechanistic and learning-based components and their interactions and to establish transparent methods for communicating these uncertainties to stakeholders to foster trust.
Finally, implementing user-centered design principles will be crucial
for enabling intuitive and effective human-digital twin interaction.

Nonetheless, the trajectory is clear. As computational science increasingly converges with cognitive computing, digital twins are evolving from analytical instruments into partners in discovery, design, and operation. The integration of agentic AI, LLMs, and AR is enabling digital twins that communicate naturally, reason autonomously, and learn continuously from both data and human interaction, all while adhering to safety, regulatory, and sustainability imperatives. Together, these developments chart a structured pathway toward more integrated cognitive systems, where perception, reasoning, and action coalesce within transparent, goal-aligned, and human-centered digital ecosystems. 




\backmatter





\bmhead{Acknowledgements}
O.S. acknowledges support from the AFOSR Grant FA9550-24-1-0327. 

\section*{Declarations} \vspace{-5pt}


\bmhead{Author contribution}
All authors contributed to the conception, writing and editing of the paper. 

\bmhead{Conflict of interest/Competing interests}
The authors declare no competing interests.

\bmhead{Ethics approval and consent to participate}
This work did not involve human participants, animal subjects, or any experiments requiring institutional ethics approval.

\end{document}